\documentclass[a4paper,11pt]{article}
\usepackage{pos}
\usepackage{graphicx}
\usepackage{wrapfig}
\usepackage{caption}
\usepackage{subcaption}

\title{Light Dark Matter in a Blazar-heated Universe}
 \ShortTitle{Light DM in Blazar-heated Universe}

\author*[a]{Oindrila Ghosh}
\author[b]{Sankalan Bhattacharyya}

\affiliation[a]{Stockholm University and The Oskar Klein Centre for Cosmoparticle Physics,\\ Alba Nova, 10691 Stockholm, Sweden}

\affiliation[b]{Cambridge University, Trinity College,\\ Cambridge CB2 1TQ, UK}

\emailAdd{oindrila.ghosh@fysik.su.se}

\abstract{Prompt emissions from TeV blazars pair produce off the extragalactic background light and the highly energetic resulting pair beams then cascade through inverse Compton scattering to give rise to secondary gamma-rays. Such reprocessed cascade emission that can be associated with individual blazar sources has not been detected thus far. The absence of pair halos around these sources, along with the non-observation of isotropic gamma-ray background excess, seems to suggest that collective plasma effects, such as beam-plasma instabilities, can play a crucial role in alleviating this GeV-TeV tension by transferring the energy from the pair beams into the background plasma of the intergalactic medium (IGM). This has profound implications not only for TeV astrophysics, but also the strength of the intergalactic magnetic field and properties of dark matter (DM). A direct consequence of the instability losses and IGM heating is the modification of thermal history at late times, which suppresses structure formation particularly in baryonically underdense regions, potentially holding a clue towards resolving the small-scale crisis in cosmology. In a blazar-heated universe, the observation of dwarf galaxies and Lyman-$\alpha$ measurements present a favoured mass range for DM candidates such as light axion-like particles.}

\ConferenceLogo{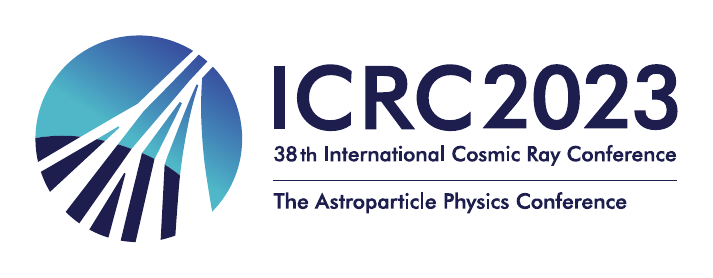}

\FullConference{%
38th International Cosmic Ray Conference (ICRC2023)\\
  26 July - 3 August, 2023\\
  Nagoya, Japan}

\begin{document}
\maketitle

\section{Introduction}

TeV blazars, ubiquitous in the gamma-ray sky, have prompt TeV emission that is reprocessed through various cascading mechanisms into the GeV band. At first, the TeV gamma rays, upon reaching the circumgalactic medium pair produce off the extragalactic background light (EBL), which is a combination of diffuse X-ray emission from star formation as well as the cosmic microwave background (CMB). The most abundantly produced pairs, electrons and positrons, can form a pair beam that propagates through large cosmological distances, well into the cosmic voids. From the injection at production, the pairs are expected to lose energy through inverse Compton scattering with the CMB photons, leading to a continuous cascade that creates a GeV flux in the 100 GeV - 1 TeV band. The non-observation of such reprocessed GeV emission \citep{aharonian2023constraints} has led to three broad classes of hints. 

\begin{itemize}
    \item A. \textit{Pair deflection:} Deflection of the charged $e^{+}e^{-}$ pairs is thought to deviate the visible flux of the cascaded gamma rays away from the line of sight. The efficacy of this mechanism depends on the strength of the intergalactic magnetic field (IGMF).
    \item B. \textit{Collective plasma effects:} When the pairs in blazar beams interact with the background thermal plasma of the intergalactic medium (IGM), unstable modes of the Langmuir excitations in the beam can grow leading to an energy drain from the pair beams into the IGM. However, the efficiency of this process is debated and is not entirely independent of the IGMF.
    \item C. \textit{Beyond the Standard Model processes:} The existence of dark matter (DM) candidates such as axion-like particles (ALPs) that can interact with photons and $e^{+}e^{-}$ may be able to eliminate some of the energy away from the cascades.
\end{itemize}

When the IGMF strength is $\sim \mathcal{O}(10^{-9}~\mathrm{G})$, the pairs are deflected sufficiently far away from the blazars such that the cascade emission derived from them cannot be associated with a specific source. This immediately implies an additional contribution to the isotropic gamma-ray background (IGRB). However, the contribution of the so-called ``known'' components to the IGRB is well-understood and primarily comes from various sources such as misaligned AGNs and star-forming galaxies, which limits the diffuse blazar contribution to $\lesssim 10\%$. The absence of any observed IGRB excess thus points towards a weaker IGMF \citep{2023arXiv230301524B}. In the instance it is weaker than nG-strength, the deflection of beam pairs should lead to the creation of a diffuse pair halo around the TeV blazars. Since such pair halos have not been observed by gamma-ray telescopes such as Fermi-LAT \citep{chen2015search}, the IGMF could be weaker than $10^{-14}~\mathrm{G}$. In a recent analysis using 1750 blazars from the Fermi 4FGL catalogue, \citep{2023arXiv230301524B} has inferred that for blazars with an intrinsic power-law cutoff above 5 TeV, the standard astrophysics of TeV blazars is bound to overproduce the IGRB. The absence of such an IGRB excess then introduces a sliding scale in the IGMF strength corresponding to a correlation length of 1 Mpc, while the magnetic deflection or diffusion competes with collective plasma effects. 

On the one hand, for an IGMF strength of  $10^{-18}-10^{-14}~\mathrm{G}$, either plasma instabilities or magnetic diffusion of the pair beams could be important, while closer to the upper limit, the magnetic diffusion can quench the instability growth. On the other hand, for an IGMF strength $\lesssim \mathcal{O}(10^{-18}~\mathrm{G})$, plasma instability is the leading explanation for the missing cascade within the Standard Model. In the subsequent sections, we outline how instability-induced heating of the IGM leads to interesting consequences for structure formation in the small scale at late times via its thermal history and indicates a preferred mass window for light DM in presence of such blazar heating.

\section{Plasma instability and IGM heating}

\subsection{Beam evolution}

Amidst various growing modes of instability, electrostatic instabilities are deemed the most efficient in cooling the cosmic pair beams; however, their exact efficiencies are debated and depend strongly on parameters such as the beam Lorentz factor, the ratio of the number density of particles in the beam to that in the background plasma, beam temperature characterised by the initial width of the momentum distribution at pair injection etc. The induced electric fluctuations within the beam, due to instabilities or inhomogeneities in the background plasma, leads to a self-heating of the beam that can subsequently increase its angular width and suppress the instability growth. This interplay between the energy dissipation and diffusion shapes the evolution of the beam-plasma system, which can be described in a compact form with a Fokker-Planck equation, as shown for the first time in \citep{beck2023evolution}:

\begin{equation}
\frac{\partial}{\partial t} f(\mathbf{p}, t)= -\frac{\partial}{\partial p_i}\left[V_i(\mathbf{p}, t) f(\mathbf{p}, t)\right]+\frac{\partial}{\partial p_i}\left[D_{ij}(\mathbf{p}, k, t) \frac{\partial}{\partial p_j} f(\mathbf{p}, t)\right]\,,
\label{eq:fp}
\end{equation}
 
where in the RHS the first drift term represents energy loss owing to instability $\mathbf{V}(\mathbf{p}, t) = \dot{\mathbf{p}}$ and the second diffusion term shows momentum diffusion through coefficient $D_{ij}$. As the electrostatic modes grow, the mode-dependent spectral energy density of the electric field $W(k)$ increases as 

\begin{equation}
    W(k,t) = W_{0} \int_0^{\tau} e^{2\delta_{i}(k) \omega_{p} t} dt
\end{equation}

over one e-folding time, i.e,  growth time $\mathcal{T} \sim 1/(\delta_{i} \omega_{p})$ with generic dimensionless growth rate $\delta_{i}$ and plasma frequency $\omega_{p}$. For an astrophysical pair beam, a Lorentz factor of $\gamma_{b} \sim 10^5 - 10^6$ and a density contrast of $\alpha \sim 10^{-18}-10^{-15}$ reflect standard conditions.

\subsection{Thermal history of the intergalactic medium}

The intergalactic medium undergoes a plethora of heating and cooling processes, the efficiency of which depends on the local overdensity. Defining the matter overdensity as $\delta = (\rho - \overline{\rho})/\overline{\rho}$, with $\rho$ being the density at a given location, and $\overline{\rho}$ the cosmic mean density, the thermal evolution of the IGM is best described as the phenomenological relation:

\begin{equation}
    T = T_0 (1+\delta)^{\gamma^{\prime}(z)-1} = T_0 \Delta^{\gamma^{\prime}(z)-1}
\end{equation}

where $\gamma^{\prime}$ denotes the index of the tight power law, often characterised as the equation of state for the IGM, $T$ is the temperature of the IGM at a given redshift $z$, and the mean density temperature profile, $T_{0}$ \citep{mcquinn2016evolution}. The thermal evolution of the IGM at late times is governed by four components: a) cooling due to Hubble expansion, b) increase in entropy due to structure formation, c) photoionization, photoheating and various other redshift-dependent processes, and d) heating of the IGM through other mechanisms \citep{mcquinn2016evolution}. The IGM temperature as a function of redshift and therefore time is then \citep{mcquinn2016intergalactic}:

\begin{equation}
\frac{d T}{d t}=-2 H T+\frac{2 T}{3 \Delta} \frac{d \Delta}{d t}+\frac{2}{3 k_B n_{\mathrm{bary}}} \frac{d Q}{d t}
\label{eq:trhoeq}
\end{equation}

where the first term on the RHS denotes Hubble cooling due to the expansion of the universe. Overdensity is represented by $\Delta$ and $n_{\mathrm{bary}}$ is the baryon number density. The blazar heating term is expressed as $\dot{Q}_{B}$, and all other standard heating and cooling processes are expressed in units of 3860 K per free particle per Gyr, which add to the third term in the RHS as $\Sigma_{\mathrm{std}}$ such that \citep{mcquinn2016evolution} 

\begin{equation}
    \dot{Q} = \dot{Q}_{\mathrm{B}} + \frac{3 k_B n_b}{2} \Sigma_{\mathrm{std}} \dot{\mathcal{Q}}
\end{equation}

where $\Sigma_{\mathrm{std}} \dot{\mathcal{Q}} = \dot{\mathcal{Q}}_{\mathrm{H-I,photo}} + \dot{\mathcal{Q}}_{\mathrm{He-I,photo}} + \dot{\mathcal{Q}}_{\mathrm{He-II,photo}} + \dot{\mathcal{Q}}_{\mathrm{H-II,rec}} + \dot{\mathcal{Q}}_{\mathrm{He-III,rec}} + \dot{\mathcal{Q}}_{\mathrm{Compton}} + \dot{\mathcal{Q}}_{\mathrm{free-free}}$. For a heating rate contribution of $\dot{Q}_{B}$ from blazars, we use the parameterisation based on 40 observed blazars within 1-$\sigma$ uncertainty as shown in \citep{chang2012cosmological}, \citep{puchwein2012lyman}:

\begin{equation}
\begin{aligned}
\log _{10}\left(\frac{\dot{Q}_{\mathrm{B}} / n_{\text {bary }}}{1 \mathrm{eVGyr}^{-1}}\right)= 0.0315(1+z)^3-0.512(1+z)^2 +2.27(1+z)-\log _{10} \dot{Q}_{\mathrm{mod}}.
\end{aligned}
\end{equation}

\begin{figure}
\centering
\includegraphics[width=.7\linewidth]{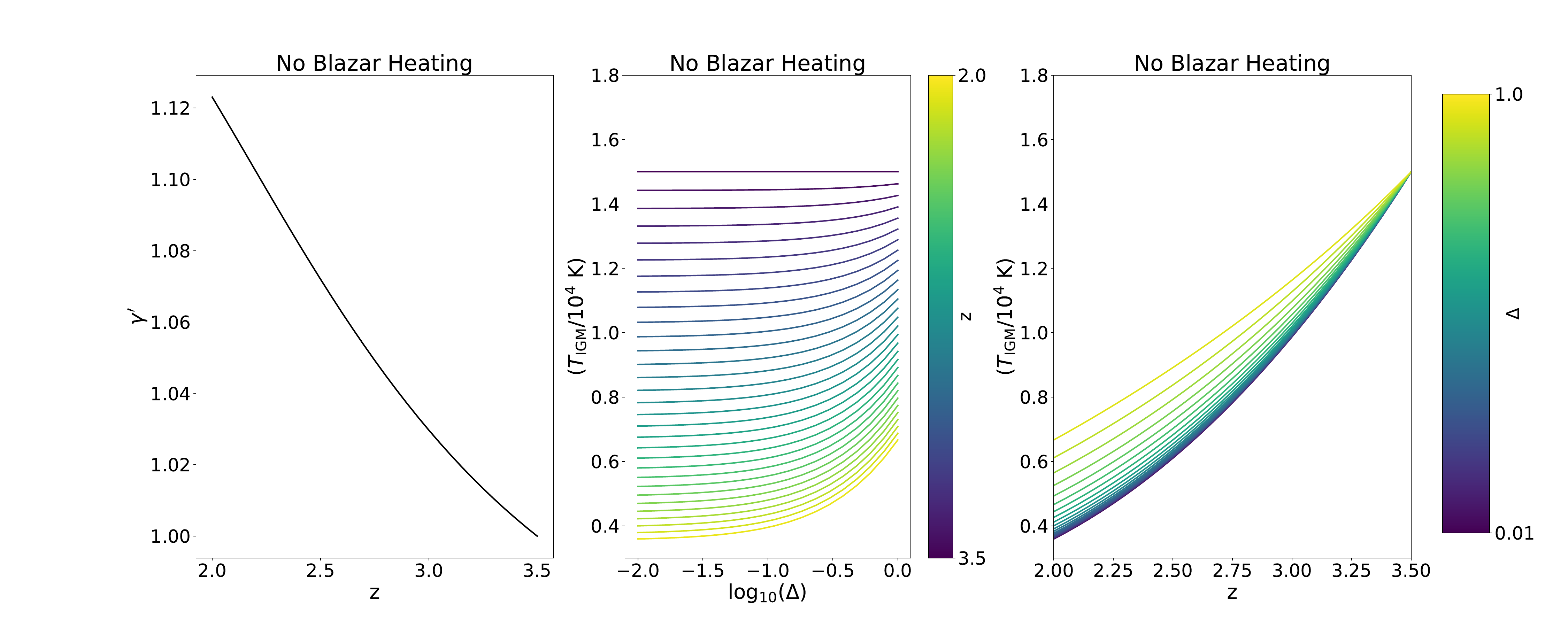}
\caption{Temperature-density-redshift relation without blazar heating}
\label{fig:trhonobl}
\end{figure}

\begin{figure}
\centering
\includegraphics[width=.7\linewidth]{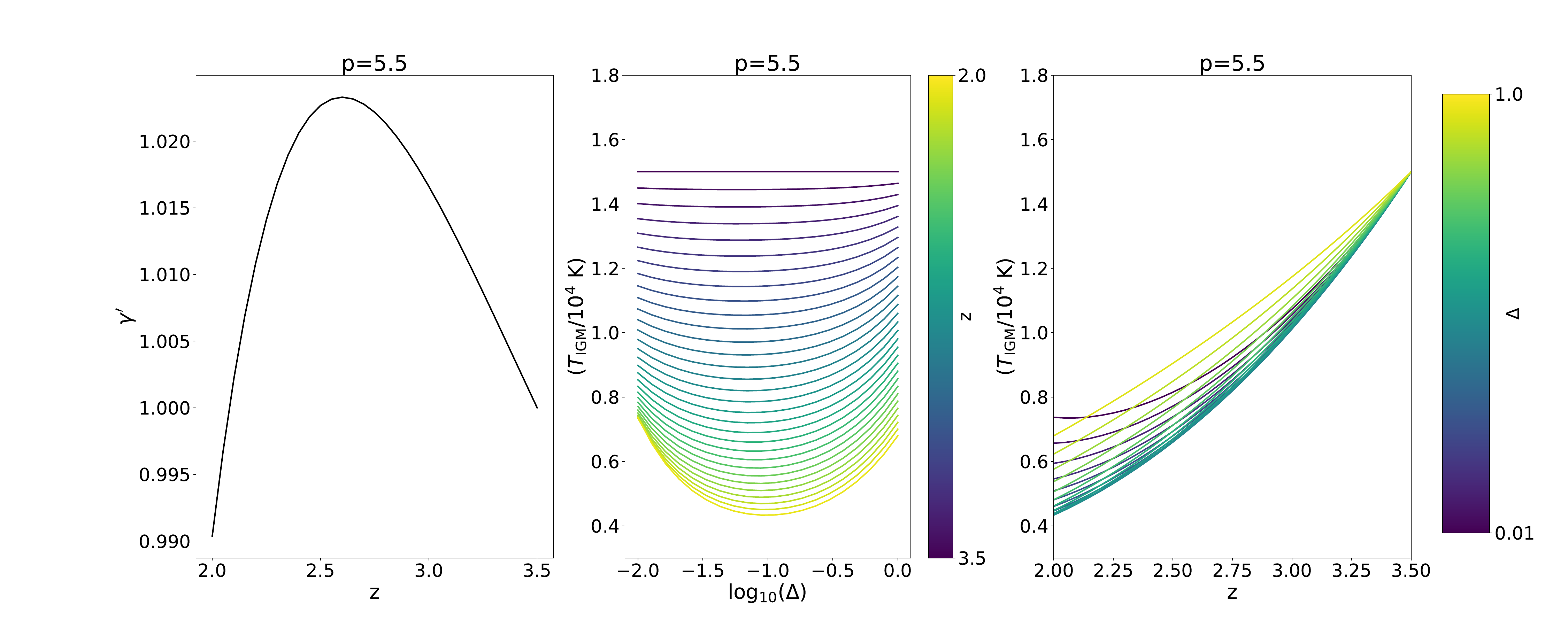}
\caption{Temperature-density-redshift relation for weak blazar heating characterised by $p=5.5$}
\label{fig:trhop5.5}
\end{figure}

\begin{figure}
\centering
\includegraphics[width=.7\linewidth]{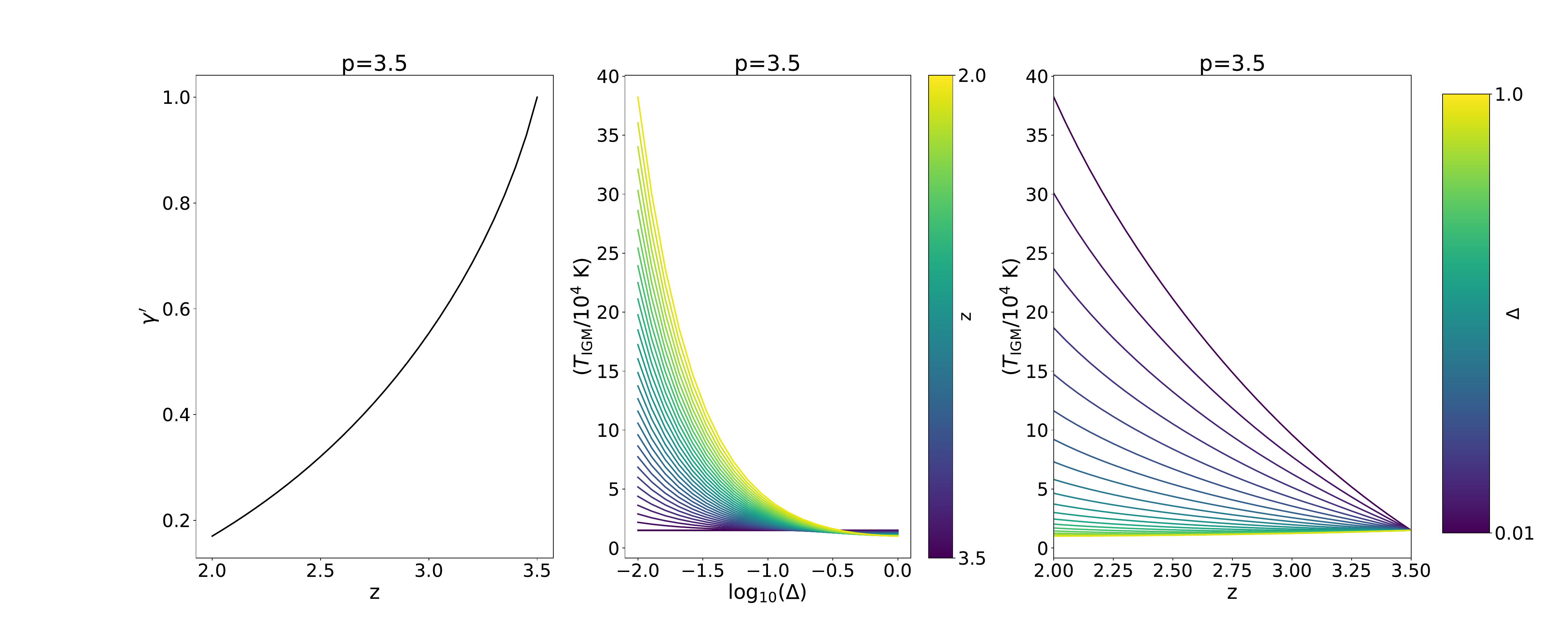}
\caption{Temperature-density-redshift relation for strong blazar heating characterised by $p=3.5$}
\label{fig:trhop3.5}
\end{figure}

\begin{wrapfigure}{R}{0.4\textwidth}
\centering
\includegraphics[width=.88\linewidth]{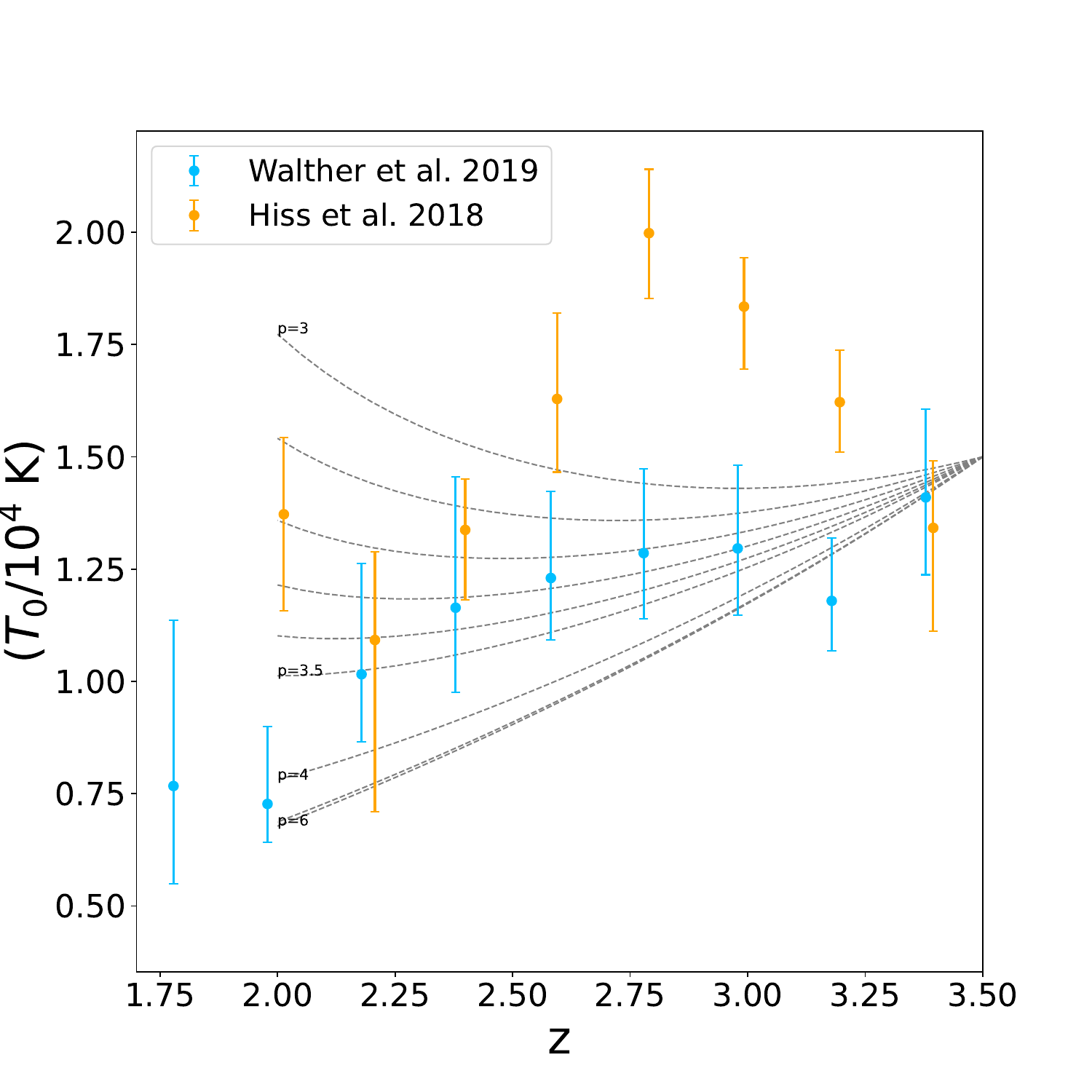}
\captionof{figure}{Temperature at mean density in units of $10^{4}~\mathrm{K}$ plotted against redshift for various degrees of blazar heating within the range $p=3-6$. Lyman-$\alpha$ measurements of the IGM temperature at mean density are shown as datapoints in orange for Hiss et al. (2018) \citep{hiss2018new} and blue for Walther et al. (2019)\citep{walther2019new}.}
\label{fig:t0zdata}
\end{wrapfigure}

In this work, the following parameter $p$ describes the contribution of blazar heating as $\log _{10} \dot{Q}_{\mathrm{mod}} = p =\{3-6\} $ for moderate to weak blazar heating cases, depending on the instability growth rate. We show the temperature-density relation as a solution to Eq. \ref{eq:trhoeq}, neglecting local changes to overdensity at the redshift window of interest $z \sim 2-3.5$, in Figs. \ref{fig:trhonobl}, \ref{fig:trhop5.5}, and \ref{fig:trhop3.5} without blazar heating, for weak blazar heating ($p =5.5$), and moderate blazar heating ($p =3.5$) respectively. In Fig. \ref{fig:trhonobl}, the slope of the temperature-density relation is qualitatively consistent with canonical heating mechanisms described above \citep{mcquinn2016evolution} without shocks or additional heating components such as plasma instabilities. Subsequently, in presence of weak blazar heating $p=5.5$, we witness the transition to inverted $T-\rho$ and $T-z$ relations in Fig. \ref{fig:trhop5.5}, while the expected inversion is clearly visible in Fig. \ref{fig:trhop3.5} for moderate blazar heating $p=3.5$. In order to understand the realistic heating scenario, we plot the IGM temperature at mean density for a range of $p$, as shown in Fig. \ref{fig:t0zdata} using Lyman-$\alpha$ datasets \citep{walther2019new},\citep{hiss2018new}.

 We have set $z=3.5$ as the end of reionization while initializing our computation in solving for the redshift-dependent temperature-density relation. The main source of uncertainty in the calculation stems from the HeII models used.

\section{Impact on late-time structure formation}

Fuelled by the discrepancies between cosmological simulations and observations of dwarf galaxies, there is a long-standing doubt cast on the validity of the collisionless $\Lambda$CDM scenario in the small-scale, in terms of the so-called ``missing satellite problem'', ``cusp vs. core issue'', and the ``too-big-to-fail problem'' \citep{bullock2017small}. Apart from modelling various baryonic feedback processes, modifications to the standard cold dark matter paradigm have been proposed by either making DM collisional (self-interacting) or relativistic to introduce a cutoff scale (warm DM), or invoking the fluid-like behaviour of light scalars such as ALPs. In presence of blazar heating, a hotter IGM leads to an elevated pressure scale, with its impact on late-forming structures of particular significance. In order to understand how this fares against the predictions of DM models that suppress the small-scale power, correction due to both to the mass confined within the corresponding pressure scale, i.e., the Jeans masses should be taken into account. 

\subsection{Modification of pressure scale}

The Jeans wave mode for a halo depends on the sound speed 
$k_{\mathrm{J, th}}(a) \equiv \left(a/c_{\mathrm{s,th}}(a)\right) \sqrt{4 \pi G \bar{\rho}(a)}$ where the scale factor $a \equiv 1/(1+z)$ and $G$ is the gravitational constant. Considering the index of the equation of state $\gamma_{\mathrm{gas}}=5/3$ for H \& He, the corresponding sound speed is estimated as $c_{\mathrm{s,th}}(a) \equiv \sqrt{5 k T(a)/ 3 \mu m_{p}}$ where mean molecular weight $\mu = 0.533$ such that $m =\mu m_{p} $, with $m_{p}$ as proton mass for standard IGM conditions. $T(a)$ is the redshift (scale factor)-dependent gas temperature and $k_{B}$ represents the Boltzmann's constant. In this work, we confine the discussion regarding DM to  light ALPs that have a non-zero contribution to the sound speed, introducing an axionic Jeans scale $k_{\mathrm{J, a}}(a) = \left(16 \pi G \bar{\rho}(a) m_{a}^{2} \right)^{1/4}$ where $m_{a}$ is the ALP mass.


Changes in pressure require finite time to impact the IGM gas distribution in presence of Hubble expansion, and the Jeans scale only takes into account the instantaneous sound speed. Taking the entire thermal history of the IGM into account, a better measure in terms of a filtering scale $k_{F}$ can be introduced such that the baryonic and DM perturbations, written as $\delta_b$ and $\delta_X$ respectively, grow according to $\delta_b=\delta_X \exp \left[-k^2 / k_F^2\right]$. Above this scale the baryonic perturbations are smoothened in linear perturbation theory, setting a characteristic mass scale. For relatively high redshifts up until after reionization $2 < z \lesssim 3.5$, we can compute the filtering-scale wave mode from an effective Jeans scale using an Einstein-deSitter approximation for matter-dominated models, with a corresponding filtering mass $M_{\mathrm{F}}(a) \equiv \left(4 \pi/3 \right) \bar{\rho}(a)\left(2 \pi a/k_F(a)\right)^3$ such that

\begin{equation}
\frac{1}{k_F^2(a)}=\frac{3}{a} \int_{a_{\min }}^a d a^{\prime} \frac{1}{k_J^2\left(a^{\prime}\right)}\left[1-\left(\frac{a^{\prime}}{a}\right)^{1 / 2}\right]
\label{eq:kfhighz}
\end{equation}

\subsection{Void dwarfs and limits on light ALP mass}

\begin{figure}[htbp!]
\centering
\begin{tabular}{c}
  \includegraphics[width=0.3\linewidth]{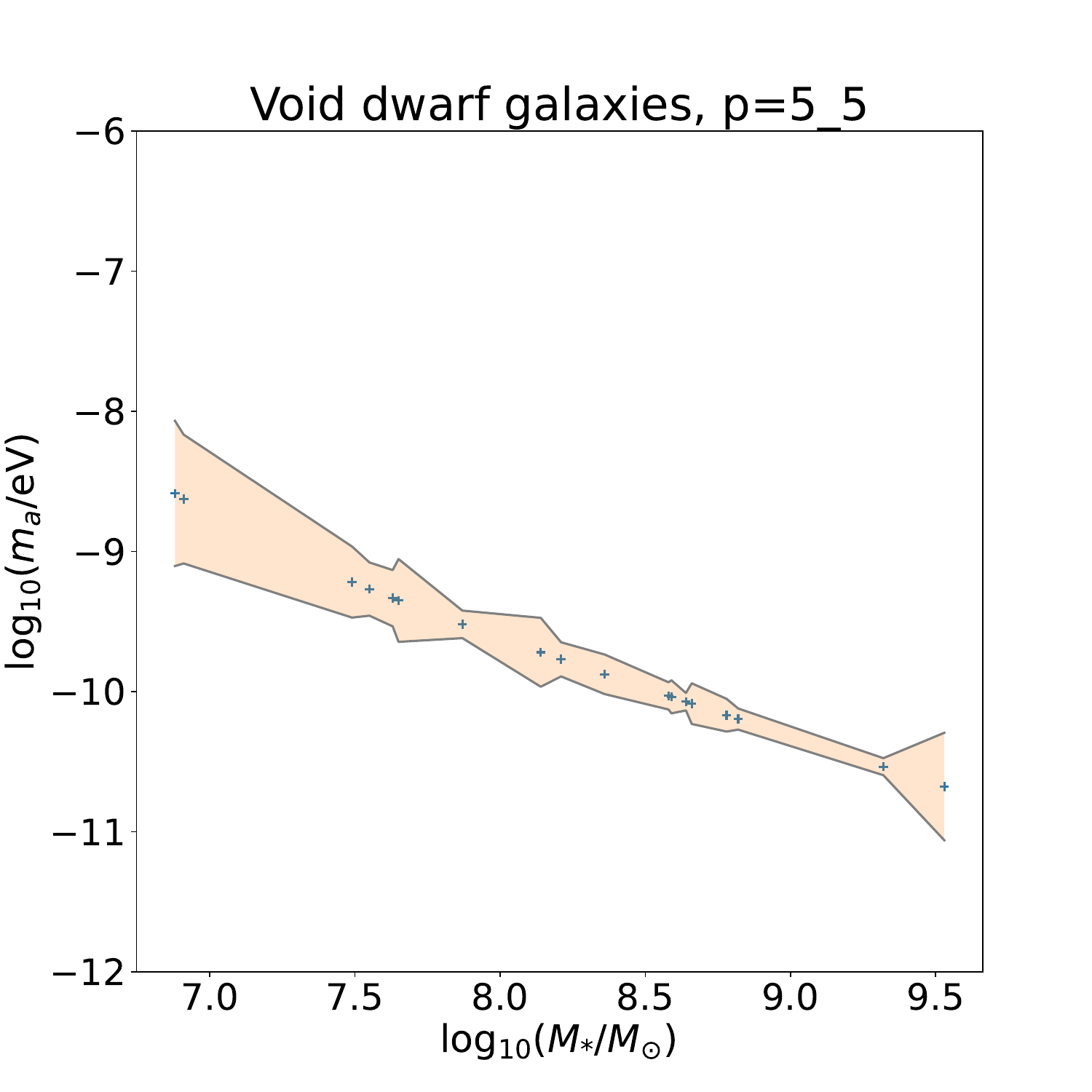}
  \small{(a)}
\end{tabular}
\hspace{0.01\linewidth}
\begin{tabular}{c}
  \includegraphics[width=0.3\linewidth]{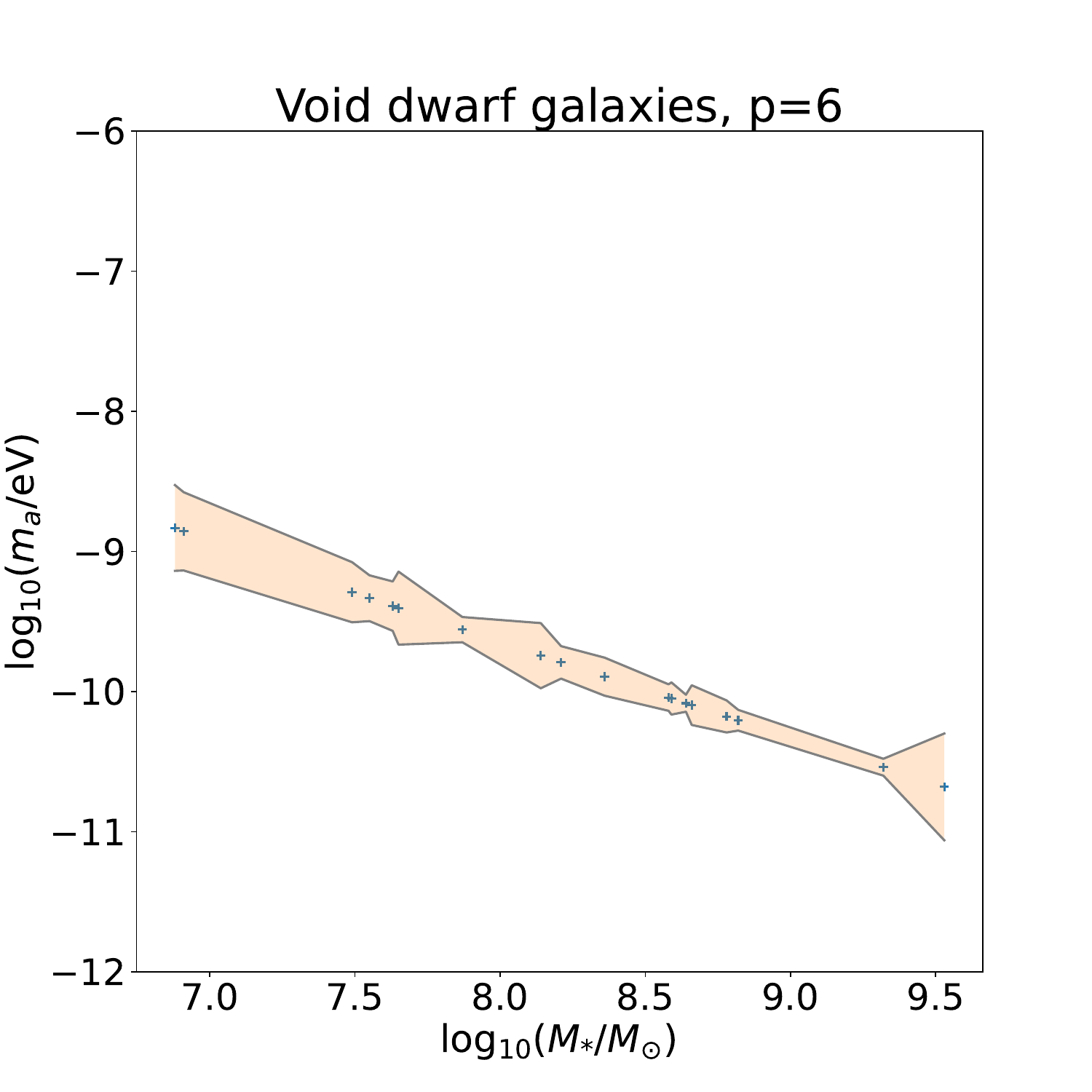}
  \small{(b)}
\end{tabular}

\begin{tabular}{c}
  \includegraphics[width=0.3\linewidth]{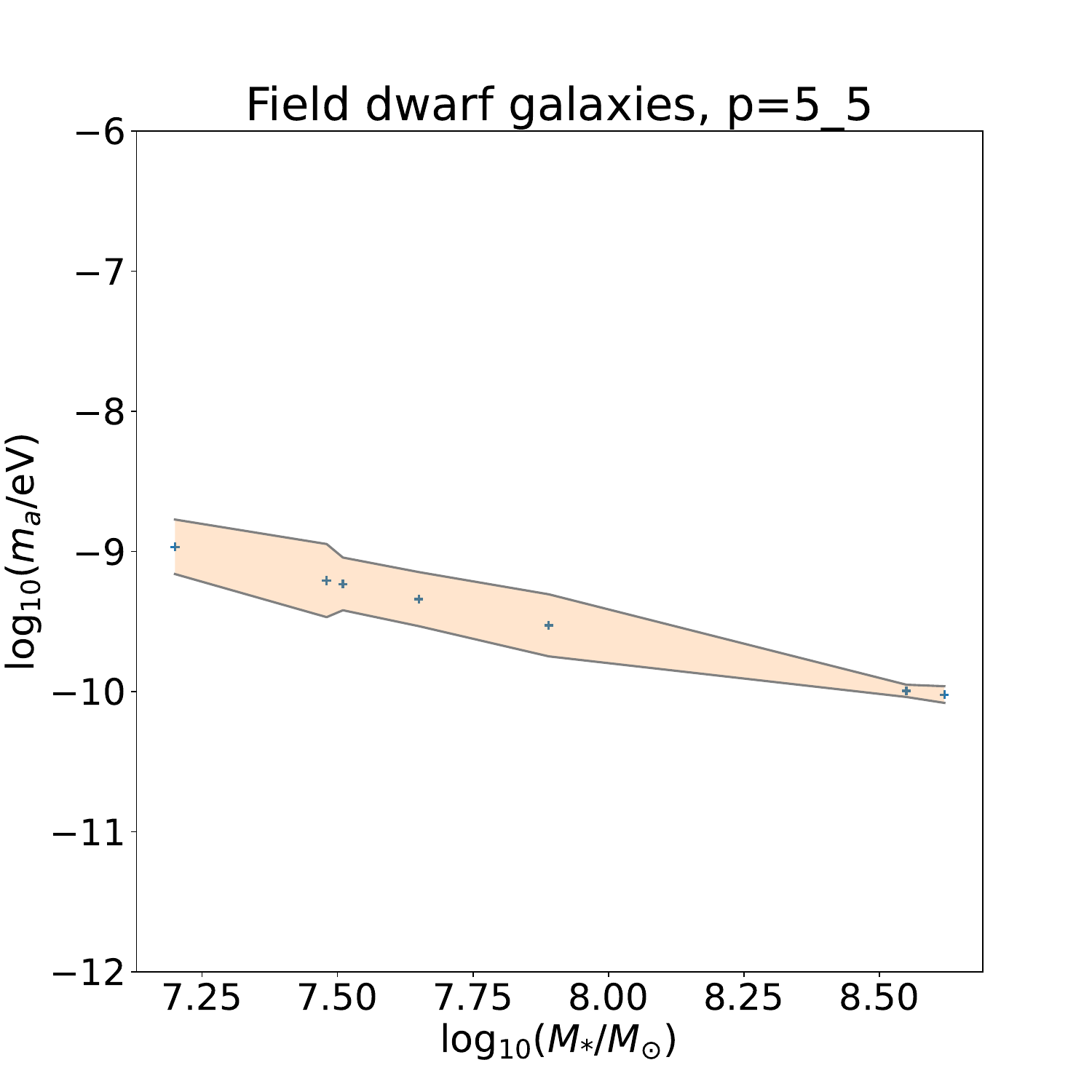}
  \small(c)
\end{tabular}
\hspace{0.01\linewidth}
\begin{tabular}{c}
  \includegraphics[width=0.3\linewidth]{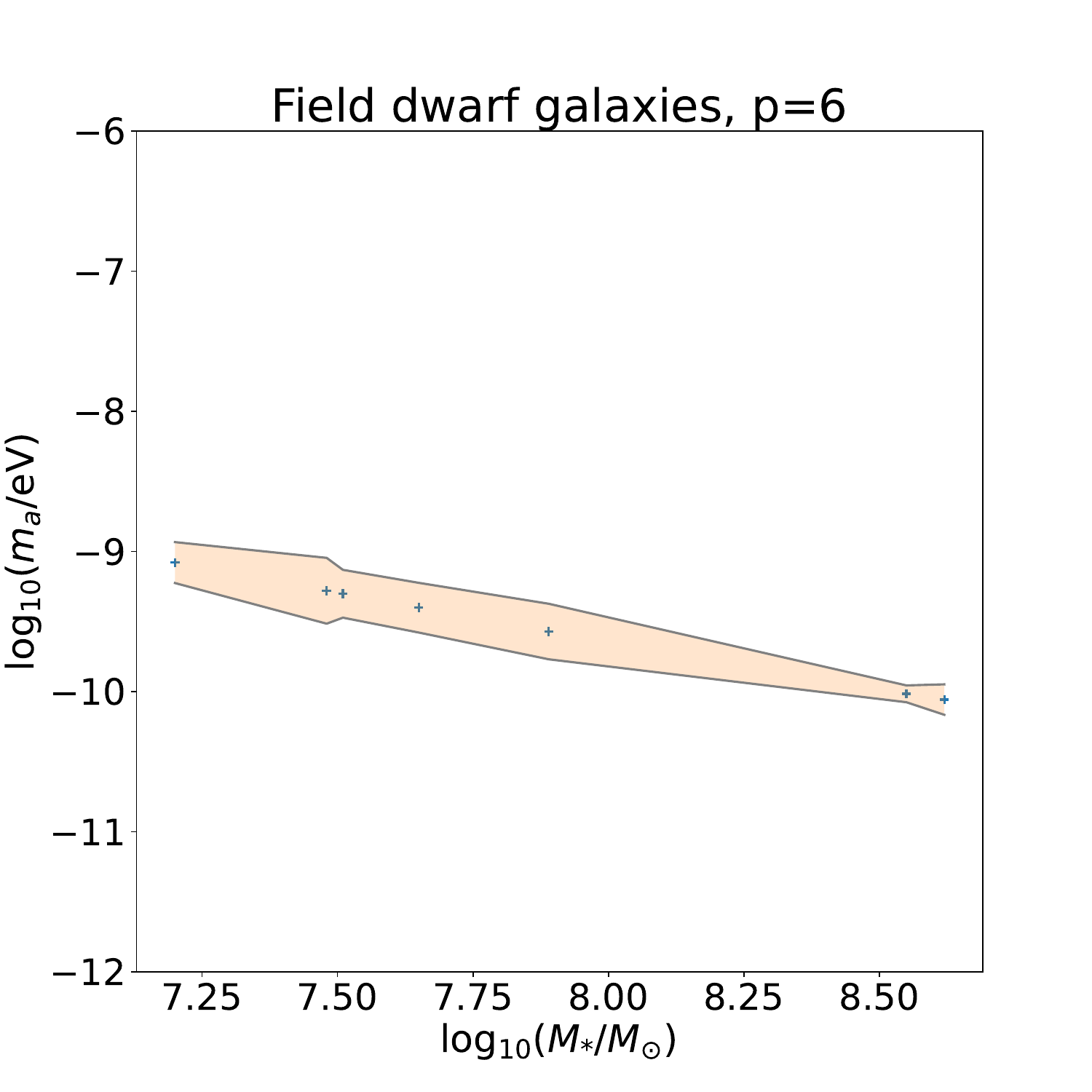}
  \small{(d)}
\end{tabular}
\caption{Favoured light ALP masses derived from KCWI measurements of dwarf galaxy characteristic masses with $1\sigma$ contour, with top panels (a)-(b) for void dwarfs and bottom panels (c)-(d) for field dwarfs. The panels on the left side (a), (c) correspond to $p=5.5$ and panels to the right (b), (d) are for $p=6$.}
\label{fig:void-field-ma}
\end{figure}

We note from Figs. \ref{fig:trhop5.5} and \ref{fig:trhop3.5} that the impact of blazar heating is most prominent in underdense regions. Dwarf galaxies are largely dark-matter dominated structures that are isolated when residing in a cosmic void, and their merger and accretion histories are thus significantly less rich compared to their counterparts in denser environments. Thus, void dwarfs serve as excellent testbeds for exploring the permitted parameter ranges for light DM in presence of blazar heating.

Characteristic masses of dwarf galaxies in the field and the local void have been estimated using the latest stellar kinematics measurements from the Keck Cosmic Web Imager (KCWI) \citep{de2023dwarfs}. We derive the corresponding ALP masses for void and field dwarfs that are presented in Fig. \ref{fig:void-field-ma} shown as blue crosses with $1\sigma$ error as the beige contour, in panel (a)-(b) and (c)-(d) respectively. The left (right) panels represent $p=5.5$ ($p=6$). We note that dwarf galaxy characteristic masses are comparable to halo filtering masses at a given redshift, especially for those without significant environmental effects such as mergers and tidal disruptions. From Fig. \ref{fig:void-field-ma}, we infer a favoured ALP mass range of $10^{-10.5}-10^{-8.5}~\mathrm{eV}$ for void dwarfs, and $10^{-10}-10^{-9}~\mathrm{eV}$ for field dwarfs, using the degree of heating parameterised as $p=5.5$ and $p=6$. For our analysis, we chose weak heating to conservatively demonstrate the impact of blazar heating on the filtering scale and the ALP masses they correspond to.

\section{Summary and outlook}

The absence of an observed IGRB excess due to diffuse blazar cascades and non-observation of pair halos associated with TeV blazars indicate a feeble IGMF that competes with plasma instabilities in resolving the GeV-TeV tension. Such collective plasma effects transport energy from the pair beams into the IGM through unstable electrostatic Langmuir oscillation modes, altering its thermal state locally with its effect most pronounced in underdense regions such as the cosmic voids. Even though particle-in-cell simulations estimate that $\lesssim 10\%$ of the beam energy could be lost through instabilities \citep{beck2023evolution}, it can still contribute to considerable IGM heating. Computing a global average heating from such instability-associated losses in blazar beams, we applied Lyman-$\alpha$ measurements to refine the ranges of degree of heating consistent with the IGM temperature at mean density. Our results are applicable to the linear regime, and hold particular significance in environments where the baryonic components co-evolve with DM. Assuming that all of DM is composed of stable light ALPs, we derived favoured ALP masses in the range $10^{-10.5}-10^{-8.5}~\mathrm{eV}$ and $10^{-10}-10^{-9}~\mathrm{eV}$ in presence of blazar heating based on respectively the characteristic masses of the void and field dwarf galaxies computed from the KCWI stellar kinematics datasets. 

\section*{Acknowledgements}

SB was supported by the DFG-DAAD Research Internship in Science and Engineering programme during the initial stage of the project. OG acknowledges funding from DFG and DAAD through the proposals ``Looking for Space Plasma Instabilities in the Lab (2021)'' and ``Thermal History of a Blazar-heated Universe (2022)'' and support of EXC 2121 Quantum Universe – 390833306. OG is supported by the European Research Council under Grant No. 742104 and by the Swedish Research Council (VR) under the grants 2018-03641 and 2019-02337. We thank Carol Ballinger for preliminary discussions and Joerg Jaeckel for useful comments.

\bibliographystyle{JHEP}
\bibliography{skeleton}

\providecommand{\href}[2]{#2}\begingroup\raggedright\begin{thebibliography}{10}

\bibitem{aharonian2023constraints}
F.~Aharonian, J.~Aschersleben, M.~Backes, V.B.~Martins, R.~Batzofin, Y.~Becherini et~al., \emph{Constraints on the intergalactic magnetic field using fermi-lat and hess blazar observations}, {\emph{The Astrophysical Journal Letters} {\bfseries 950} (2023) L16}.

\bibitem{2023arXiv230301524B}
C.~{Blanco}, O.~{Ghosh}, S.~{Jacobsen} and T.~{Linden}, \emph{{Where are the Cascades from Blazar Jets? An Emerging Tension in the $\gamma$-ray sky}}, \href{https://doi.org/10.48550/arXiv.2303.01524}{\emph{arXiv e-prints} (2023) arXiv:2303.01524} [\href{https://arxiv.org/abs/2303.01524}{{\ttfamily 2303.01524}}].

\bibitem{chen2015search}
W.~Chen, J.H.~Buckley and F.~Ferrer, \emph{Search for gev $\gamma$-ray pair halos around low redshift blazars}, {\emph{Physical review letters} {\bfseries 115} (2015) 211103}.

\bibitem{beck2023evolution}
M.~Beck, O.~Ghosh, F.~Gr{\"u}ner, M.~Pohl, C.B.~Schroeder, G.~Sigl et~al., \emph{Evolution of relativistic pair beams: Implications for laboratory and tev astrophysics}, {\emph{arXiv preprint arXiv:2306.16839} (2023) }.

\bibitem{mcquinn2016evolution}
M.~McQuinn, \emph{The evolution of the intergalactic medium}, {\emph{Annual Review of Astronomy and Astrophysics} {\bfseries 54} (2016) 313}.

\bibitem{mcquinn2016intergalactic}
M.~McQuinn and P.R.~Upton~Sanderbeck, \emph{On the intergalactic temperature--density relation}, {\emph{Monthly Notices of the Royal Astronomical Society} {\bfseries 456} (2016) 47}.

\bibitem{chang2012cosmological}
P.~Chang, A.E.~Broderick and C.~Pfrommer, \emph{The cosmological impact of luminous tev blazars. ii. rewriting the thermal history of the intergalactic medium}, {\emph{The Astrophysical Journal} {\bfseries 752} (2012) 23}.

\bibitem{puchwein2012lyman}
E.~Puchwein, C.~Pfrommer, V.~Springel, A.E.~Broderick and P.~Chang, \emph{The lyman $\alpha$ forest in a blazar-heated universe}, {\emph{Monthly Notices of the Royal Astronomical Society} {\bfseries 423} (2012) 149}.

\bibitem{walther2019new}
M.~Walther, J.~O{\~n}orbe, J.F.~Hennawi and Z.~Luki{\'c}, \emph{New constraints on igm thermal evolution from the ly$\alpha$ forest power spectrum}, {\emph{The Astrophysical Journal} {\bfseries 872} (2019) 13}.

\bibitem{hiss2018new}
H.~Hiss, M.~Walther, J.F.~Hennawi, J.~O{\~n}orbe, J.M.~O’Meara, A.~Rorai et~al., \emph{A new measurement of the temperature--density relation of the igm from voigt profile fitting}, {\emph{The Astrophysical Journal} {\bfseries 865} (2018) 42}.

\bibitem{bullock2017small}
J.S.~Bullock and M.~Boylan-Kolchin, \emph{Small-scale challenges to the $\lambda$ cdm paradigm}, {\emph{Annual Review of Astronomy and Astrophysics} {\bfseries 55} (2017) 343}.

\bibitem{de2023dwarfs}
M.A.~de~los Reyes, E.N.~Kirby, Z.~Zhuang, C.C.~Steidel, Y.~Chen and C.~Wheeler, \emph{Dwarfs in void environments (dive): The stellar kinematics of void dwarf galaxies using the keck cosmic web imager}, {\emph{The Astrophysical Journal} {\bfseries 951} (2023) 52}.

\end{thebibliography}\endgroup




%
%
%

\end{document}